\documentclass[12pt,a4paper,notitlepage,fleqn]{article}

\usepackage{setspace}
\usepackage[margin=1in]{geometry}
\usepackage[symbol]{footmisc}

\usepackage[T1]{fontenc}
\usepackage[utf8]{inputenc}
\usepackage{authblk}

\makeatletter

\newcommand{\fboxsubsec}[1]{
	\begin{flushleft}
		#1
	\end{flushleft}
	}
\renewcommand{\subsection}{\@startsection{subsection}{2}{0pt}
	{1ex}
	{0.5ex}
	{\reset@font\it\fboxsubsec}
	}
\makeatother

\title{A Test of the Adaptive Market Hypothesis using a Time-Varying AR Model in Japan}%

\author{Akihiko Noda$^{a,b}$\thanks{\scriptsize Corresponding Author. E-mail: noda@cc.kyoto-su.ac.jp, Tel. +81-75-705-1510, Fax. +81-75-705-3227.}

{\scriptsize ${}^{a}$ \it Faculty of Economics, Kyoto Sangyo University, Motoyama, Kamigamo, Kita-ku, Kyoto 603-8555, Japan}

{\scriptsize ${}^{b}$ \it Keio Economic Observatory, Keio University, 2-15-45 Mita, Minato-ku, Tokyo 108-8345, Japan}}

\date{\empty}

\renewcommand\thefootnote{\arabic{footnote}}

\usepackage[dvipdfmx]{graphicx}

\usepackage[]{natbib}%
\usepackage{amsmath,amssymb}%
\usepackage{ascmac}%
\usepackage{multirow}%
\usepackage{lscape}%
\usepackage{subfigmat}

\usepackage{pifont}%
\usepackage{arydshln}%
\usepackage[format=hang]{caption}
\usepackage[all]{xy}
\usepackage{url}
\bibpunct{(}{)}{;}{a}{}{,}

\def\hsymbu#1{\smash{\lower1.7ex\hbox{\huge$#1$}}}

\newcommand{\citetapos}[1]{\citeauthor{#1}'s \citeyearpar{#1}}

\begin{document}

\begin{titlepage}

\renewcommand{\thepage}{}
\renewcommand{\thefootnote}{\fnsymbol{footnote}}

\maketitle

\vspace{-10mm}

\noindent
\hrulefill

\noindent
{\bfseries Abstract:} This study examines the adaptive market hypothesis (AMH) in Japanese stock markets (TOPIX and TSE2). In particular, we measure the degree of market efficiency by using a time-varying model approach. The empirical results show that (1) the degree of market efficiency changes over time in the two markets, (2) the level of market efficiency of the TSE2 is lower than that of the TOPIX in most periods, and (3) the market efficiency of the TOPIX has evolved, but that of the TSE2 has not. We conclude that the results support the AMH for the more qualified stock market in Japan.\\



\noindent
\hrulefill

\end{titlepage}

\bibliographystyle{asa}


\section{Introduction}\label{sec:amhjp_intro}

Economists have recently attempted to reconcile \citetapos{fama1970ecm} efficient market hypothesis (EMH) and explore the possibility that both stock markets evolve and market efficiency varies over time (see \citetapos{lim2011esm} survey paper for more details). \citet{lo2004amh} proposes an alternative to the EMH termed the adaptive market hypothesis (AMH), which is based on an evolutionary approach. The AMH can help explain why the degree of market efficiency (or return predictability) changes over time. The most important implication of the AMH is that market efficiency can arise time to time due to changing market conditions. Therefore, a number of recent studies of the AMH have aimed to explain time variation in the degree of market efficiency.

Two approaches are used to examine the AMH. The first measures the degree of market efficiency using a time-varying model approach (see \citet{ito2014ism,ito2016eme}). They conclude that the degree of market efficiency varies over time in international stock markets. The second approach investigates market efficiency using statistical tests under the moving window method (see, \citet{kim2011srp}, \citet{lim2013aus}). \citet{kim2011srp} examine the AMH using three test statistics, namely \citetapos{choi1999trw} automatic variance ratio test, \citetapos{escanciano2009apt} automatic portmanteau test, and \citetapos{escanciano2006gst} generalized spectral test. However, the moving window method cannot avoid the empirical problem of choosing an optimal window width for these test statistics. As far as we know, only a time-varying model approach of \citet{ito2014ism,ito2016eme} can solve such an empirical problem.\footnote{Some studies have calculated the time-varying autocorrelation coefficients of stock returns as the degree of market efficiency, such as \citet{emerson1997eme}, \citet{zalewska1999efs}, and \citet{ito2009mdt}. However, their degree does not provide statistical inferences as to whether stock markets are efficient.}

This study examines \citetapos{lo2004amh} AMH in Japanese stock markets, namely the first and second sections of the Tokyo Stock Exchange, from the point of view of market efficiency. The focus of study is how their degrees of market efficiency differ from each other according to trading volume and market liberalization. Then, we measure their degrees of market efficiency using a time-varying model approach of \citet{ito2014ism,ito2016eme}. Finally, we investigate whether their degrees change over time and whether the two markets show different patterns of dynamic market efficiency depending on their trading volume and market liberalization.

\section{Hypotheses and Estimation Methods}\label{sec:amhjp_hem}

\subsection{The EMH and The AMH}\label{subsec:IRLR}
According to the EMH, market price reflects any exogenous shock at once in financial markets. Mathematically, one often represents it in the following way:
\begin{equation}
E[x_t \, | \, {\cal I}_{t-1}]=0,
\end{equation}
where $x_t$ denotes the return of a security at $t$ and ${\cal I}_{t-1}$ is the (increasing) information set at $t-1$, some $\sigma$-field to which $x_{t-1},x_{t-2},\cdots$ is adapted. Note that the EMH holds when the (log) price of the security follows a random walk process. In other words, one can say that the security price is ``determined by chance.'' 

Thus, when one considers the situation where the hypothesis does not always hold, it is natural to consider that the (excess) stock return follows a moving average process with infinite terms, $\mbox{MA}(\infty)$:
\begin{equation}
\label{eq:MA}
x_{t} = u_{t} + \beta_1 u_{t-1} + \beta_2 u_{t-2} + \cdots,
\end{equation}
where $\{u_{t}\}$ is an i.i.d. process. Since ${\cal I}_{t-1}$ is a $\sigma$-field to which $x_{t-1}$ is adapted and $\{{\cal I}_{t-1}\}$ a system of increasing information sets, the following equation holds:
\begin{equation}
E[x_t \, | \, {\cal I}_{t-1}] = \beta_1 u_{t-1} + \beta_2 u_{t-2} + \cdots. \nonumber
\end{equation}
Then, the EMH holds if and only if $\beta_i = 0$ for all $i$. 

\citet{lo2004amh} proposes the AMH, which is based on an evolutionary approach to economic interactions. He calculates the time-varying first-order autocorrelations by using the moving window method and shows that efficient and inefficient periods exist in stock markets. However, the time-varying structure of stock market efficiency remains to be elucidated. We consider the AMH that the degree of market efficiency fluctuates over time and reflects evolving market conditions: bubbles, market crashes, legal reforms, deregulations, and technological innovations. We then measure the time-varying degree of market efficiency and investigate whether the stock market evolves over time toward efficiency. 

\subsection{A Time-Varying Model Approach}\label{subsec: TV-AR}
A time-varying model approach of \citet{ito2014ism,ito2016eme} is used to analyze financial data of which the data-generating process is time-varying. In financial economics, AR models,
\begin{equation*}
x_t = \alpha_0 + \alpha_1 x_{t-1} + \cdots + \alpha_q x_{t-q} + u_t,
\end{equation*}
have been frequently used to analyze the time series of the stock returns of a financial asset, where $\{u_t\}$ satisfies $E[u_t]=0$, $E[u^2_t]=0$, and $E[u_t u_{t-m}]=0 \ \mbox{for all} \ m$. Whereas $\alpha_\ell$'s are assumed to be constant in ordinary time series analysis, we suppose that the coefficients of AR models vary over time and apply them to real financial markets, which have experienced many financial crises such as the recent collapse of Lehman Brothers, suggesting the existence of structural changes in stock markets.
\begin{equation}
x_t = \alpha_{0,t} + \alpha_{1,t} x_{t-1} + \cdots + \alpha_{q,t} x_{t-q} + u_t, \label{obseq}
\end{equation}
where $\{u_t\}$ satisfies $E[u_t]=0$, $E[u^2_t]=0$, and $E[u_t u_{t-m}]=0 \ \mbox{for all} \ m$. We call this model a time-varying autoregressive (TV-AR) model. We further suppose that parameter dynamics restrict the parameters when we estimate a TV-AR model using data. Specifically,
\begin{equation}
\alpha_{\ell,t} = \alpha_{\ell,t-1} + v_{\ell,t}, \ (\ell=1,2,\cdots,q), \label{steq}
\end{equation}
where $\{v_{\ell,t}\}$ satisfies $E[v_{\ell,t}]=0$, $E[v^2_{\ell,t}]=0$
and $E[v_{\ell,t} v_{\ell,t-m}]=0 \ \mbox{for all} \ m \ \mbox{and} \ \ell$. We regard Equations (\ref{obseq}) and (\ref{steq}) as a system of simultaneous equations.

This model estimation has two major advantages over the conventional Bayesian method (e.g., Kalman filtering and smoothing) as follows. First, our method is quite simple and fast. Unlike the conventional Bayesian method, no iteration is required. Second, our TV-AR model is non-Bayesian because it does not necessitate the prior distributions of parameters. It implies that we can employ conventional statistical inferences (e.g., residual-based bootstrap method) on the time-varying estimates.

\subsection{Time-Varying Degree of Market Efficiency}
We next calculate the time-varying impulse responses from a TV-AR coefficients in each period, estimated by using the method described in the previous subsections; we also calculate the confidence intervals for each coefficient based on the covariance matrix estimated at the same time. While the concept of a TV-AR model is simple, two points should be made here. First, the estimated model is only an approximation of the real data-generating process, which is supposed to be a complex nonstationary process. Second, we consider the estimated $\mbox{AR}(q)$ model index by period $t$, which is stationary, as a local approximation of the underlying complex process.

We define the time-varying degree of market efficiency as a special case of \citetapos{ito2014ism} one. In practice,
\begin{equation}
 \zeta_t=\left|\frac{\sum_{j=1}^p\hat{\alpha}_{j,t}}{1-\left(\sum_{j=1}^p\hat{\alpha}_{j,t}\right)}\right|.\label{degree}
\end{equation}
Note that this degree measures the deviation from the zero coefficients of the corresponding TV-MA model to our TV-AR model. Hence, we find that the large deviations of $\zeta_t$ from zero to be evidence of market inefficiency. 

The degree of market efficiency $\zeta_t$ crucially depends on the sampling errors. Thus, we construct the confidence band for possible $\zeta_t$'s on the condition that the market is efficient. We regard the market at time $t$ as inefficient whenever $\zeta_t$ is larger than the upper limit at $t$ of the band. In practice, the band is constructed as follows. First, we identify the stock returns data with the residuals of a AR($q$) estimation under the above hypothesis that all coefficients are zero. Then, we extract $N$ samples regarding it as an empirical distribution of the residuals. Secondly, we fit a TV-AR model to the $N$ bootstrap samples and derive $N$ sets of their estimates. Thirdly, we compute the $N$ bootstrap samples of $\zeta_t$ from the estimates. Finally, we construct confidence bands from the $N$ bootstrap samples (see the online appendix A.5 of \citet{ito2014ism}). That is, the bootstrap is conducted under the null hypothesis of zero autocorrelations. Hence, the estimate of the degree of efficiency outside the 99\% confidence band in Figure \ref{amhjp_fig1} means rejection of the null hypothesis of no return autocorrelation at 1\% level of significance.

\section{Data}\label{sec:amhjp_dat}

This study utilizes the monthly returns for the Tokyo Stock Price Index (TOPIX) and the Tokyo Stock Exchange Second Section Stock Price Index (TSE2) from October 1961 to December 2015, obtained from {\it the monthly statistics report of the Tokyo Stock Exchange}.\footnote{The Tokyo Stock Exchange defines the TSE2 as a free-float-adjusted market capitalization-weighted index calculated based on all the domestic common stocks listed on the Tokyo Stock Exchange Second Section.} In practice, we take the log first difference of the time series of the stock price index to obtain the returns for the TOPIX and TSE2. Table \ref{amhjp_table1} provides some descriptive statistics. We can confirm that the mean (standard deviation) of returns on the TOPIX is lower (higher) than those of the TSE2. In other words, the TSE2 is a riskier market than the TOPIX.
\begin{center}
(Table \ref{amhjp_table1} around here)
\end{center}
For the estimations, each variable that appears in the moment conditions should be stationary. To check whether the variables satisfy the stationarity condition, we use the ADF-GLS test of \citet{elliott1996eta}. Table \ref{amhjp_table1} also provides the results of the ADF-GLS test. The ADF-GLS test rejects the null hypothesis that the variables contain a unit root at conventional significance levels.\footnote{We confirm that there are no size distortions that \citet{elliott1996eta} and \citet{ng2001lls} point out in making the ADF-GLS test for small samples (see column $\hat\phi$ of Table \ref{amhjp_table1} for more details). Therefore, we use the modified Bayesian information criteria (MBIC), not the modified Akaike information criteria, to select an optimal lag order for the ADF-GLS tests.}This study utilizes the monthly returns for the Tokyo Stock Price Index (TOPIX) and the Tokyo Stock Exchange Second Section Stock Price Index (TSE2) from October 1961 to December 2015, obtained from {\it the monthly statistics report of the Tokyo Stock Exchange}.\footnote{The Tokyo Stock Exchange defines the TSE2 as a free-float-adjusted market capitalization-weighted index calculated based on all the domestic common stocks listed on the Tokyo Stock Exchange Second Section.} In practice, we take the log first difference of the time series of the stock price index to obtain the returns for the TOPIX and TSE2. Table \ref{amhjp_table1} provides some descriptive statistics. We can confirm that the mean (standard deviation) of returns on the TOPIX is lower (higher) than those of the TSE2. In other words, the TSE2 is a riskier market than the TOPIX.
\begin{center}
(Table \ref{amhjp_table1} around here)
\end{center}
For the estimations, each variable that appears in the moment conditions should be stationary. To check whether the variables satisfy the stationarity condition, we use the ADF-GLS test of \citet{elliott1996eta}. Table \ref{amhjp_table1} also provides the results of the ADF-GLS test. The ADF-GLS test rejects the null hypothesis that the variables contain a unit root at conventional significance levels.\footnote{We confirm that there are no size distortions that \citet{elliott1996eta} and \citet{ng2001lls} point out in making the ADF-GLS test for small samples. Therefore, we use the modified Bayesian information criteria (MBIC), not the modified Akaike information criteria, to select an optimal lag order for the ADF-GLS tests.}

\section{Empirical Results}\label{sec:amhjp_emp}

\subsection{Preliminary Estimations}
We assume a model with constants and use the SBIC of \citet{schwarz1978edm} as the optimal lag order selection criteria in the AR($q$) estimation. In our estimations, we choose first-order autoregressive (AR(1)) models for both the TOPIX and the TSE2. Table \ref{amhjp_table2} shows the preliminary results for the above models using the whole sample.
\begin{center}
(Table \ref{amhjp_table2} around here)
\end{center}
The AR estimates are statistically significant at conventional levels except for the constant terms in the equations. In particular, the AR(1) estimates are relatively high, about $0.3$ (TOPIX) and $0.4$ (TSE2), indicating that a shock in any month affects the return of two months later by at least 9\% (TOPIX) and 16\% (TSE2).

Now, we investigate whether the parameters are constant in the above AR(1) models using \citetapos{hansen1992a} test under the random parameters hypothesis. Table \ref{amhjp_table2} also presents the result of this parameter constancy test; we reject the null of constant parameters against the parameter variation as a random walk at the 1\% significance level. Therefore, we estimate the time-varying parameters of the above AR models to investigate whether gradual changes occur in the Japanese stock market.

\subsection{Time-Varying Degree of Market Efficiency}
Next, we employ a time-varying model approach of \citet{ito2014ism,ito2016eme} to estimate the degree of market efficiency. Since this degree is based on the spectral norm, we measure the stock markets' deviation from the efficient condition by using Equation (\ref{degree}). For example, considering the TOPIX, the degree of market efficiency tells us how the market is different from the efficient market. If $\zeta_t=0$ for time $t$, the market is shown to be efficient at that time.

\begin{figure}[!hbp]
 \caption{The Time-Varying Degree of Market Efficiency}
 \label{amhjp_fig1}
 \begin{center}
 \includegraphics[scale=0.4]{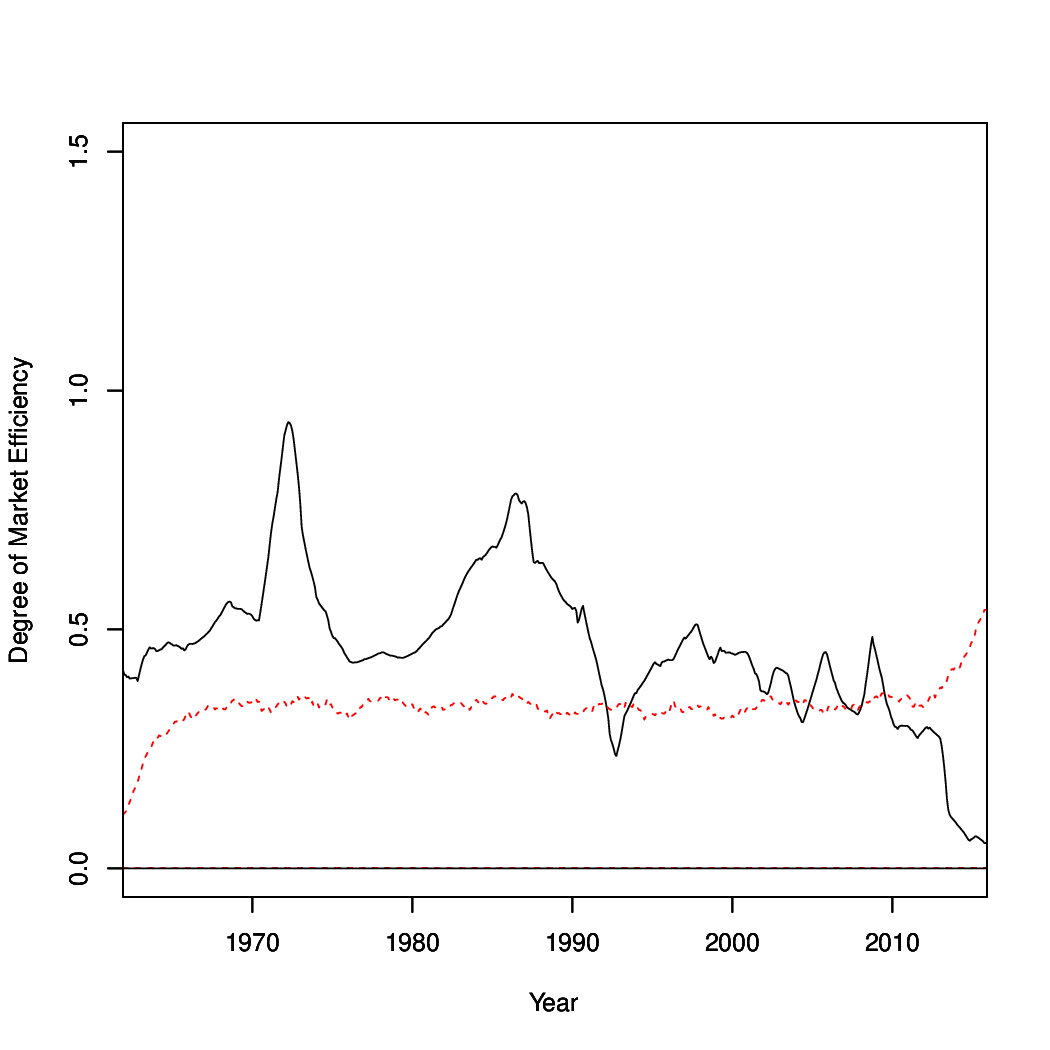}
 \includegraphics[scale=0.4]{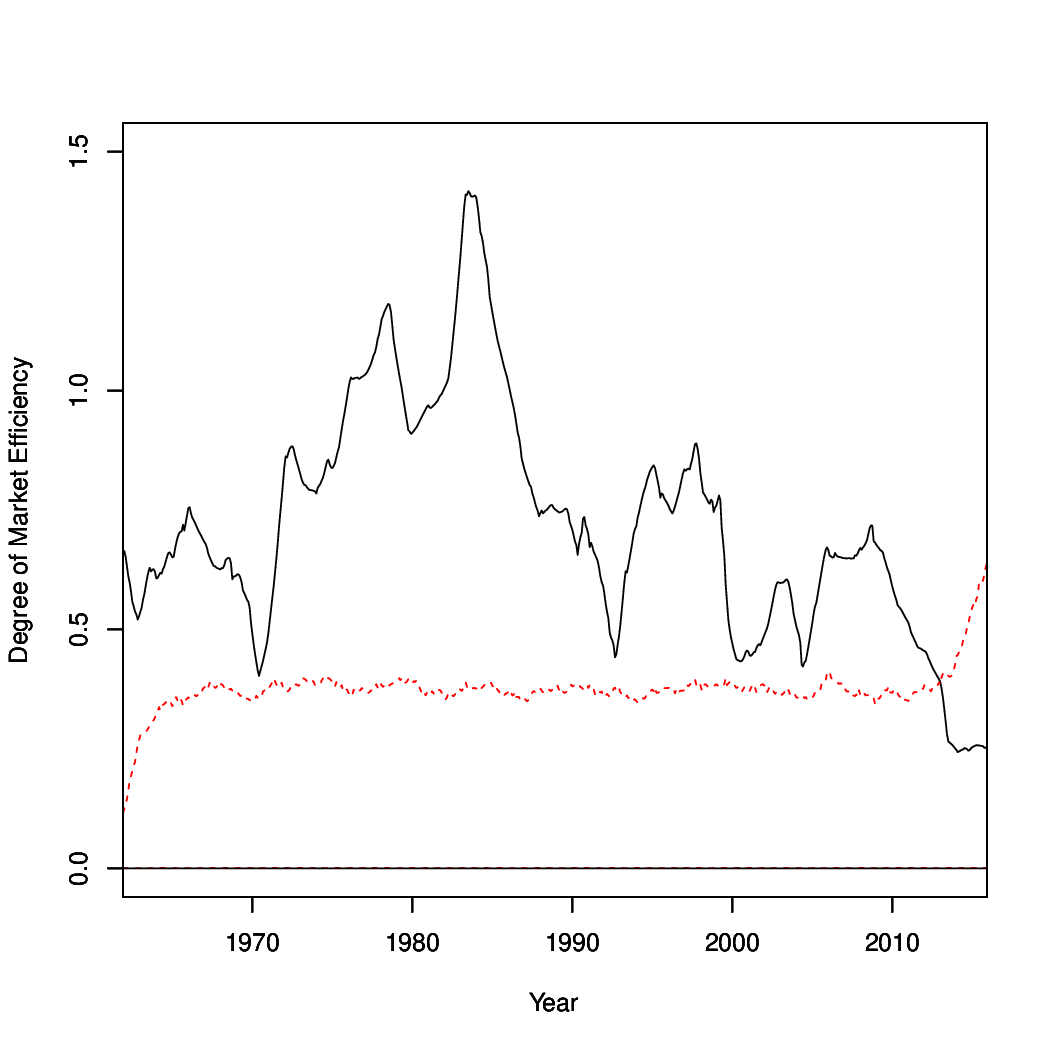}
\vspace*{5pt}
{
\begin{minipage}{420pt}
\scriptsize
\underline{Notes}:
\begin{itemize}
 \item[(1)] The panels of the figure show the time-varying degree of
	    market efficiency for the TOPIX (left panel) and the TSE2
	    (right panel).
 \item[(2)] The dashed red lines represent the 99\% confidence bands of
	    the degrees in the case of an efficient market.
 \item[(3)] R version 3.2.3 was used to compute the estimates.
\end{itemize}
\end{minipage}}%
\end{center}
\end{figure}

Figure \ref{amhjp_fig1} shows the degrees of market efficiency based on the above TV-AR(1) models.\footnote{We confirm that the models hold local stationarity by checking whether all the absolute values of the eigenvalues of each local AR(1) are less than one.} We first find that the degrees of the TOPIX and TSE2 change over time. Figure \ref{amhjp_fig1} also demonstrates the markets were completely inefficient in the 1970s and 1980s. Interestingly, these correspond with the oil crisis in the 1970s and the asset price bubble in Japan in the 1980s.
\begin{center}
(Figure \ref{amhjp_fig1} around here)
\end{center}
We confirm three significant differences between the TOPIX and TSE2 in terms of the degree of market efficiency. First, the market efficiency of the TSE2 is lower than that of the TOPIX in most periods (the averages of the TOPIX and TSE2 are about 0.46 and 0.72, respectively). Second, the market efficiency of the TSE2 fluctuates more widely than that of the TOPIX. In fact, the standard deviations of the degrees of the TOPIX and TSE2 are about 0.16 and 0.24, respectively. Third, the market efficiency of the TOPIX has been less volatile since the bursting of the bubble economy in March 1991, but that of the TSE2 has not.\footnote{See the index of business conditions by the Cabinet Office, Government of Japan (\url{http://www.esri.cao.go.jp/en/stat/di/di-e.html}).}

\begin{figure}[!hbp]
 \caption{Trading Volumes and Market Capitalizations}
 \label{amhjp_fig2}
 \begin{center}
 \includegraphics[scale=0.4]{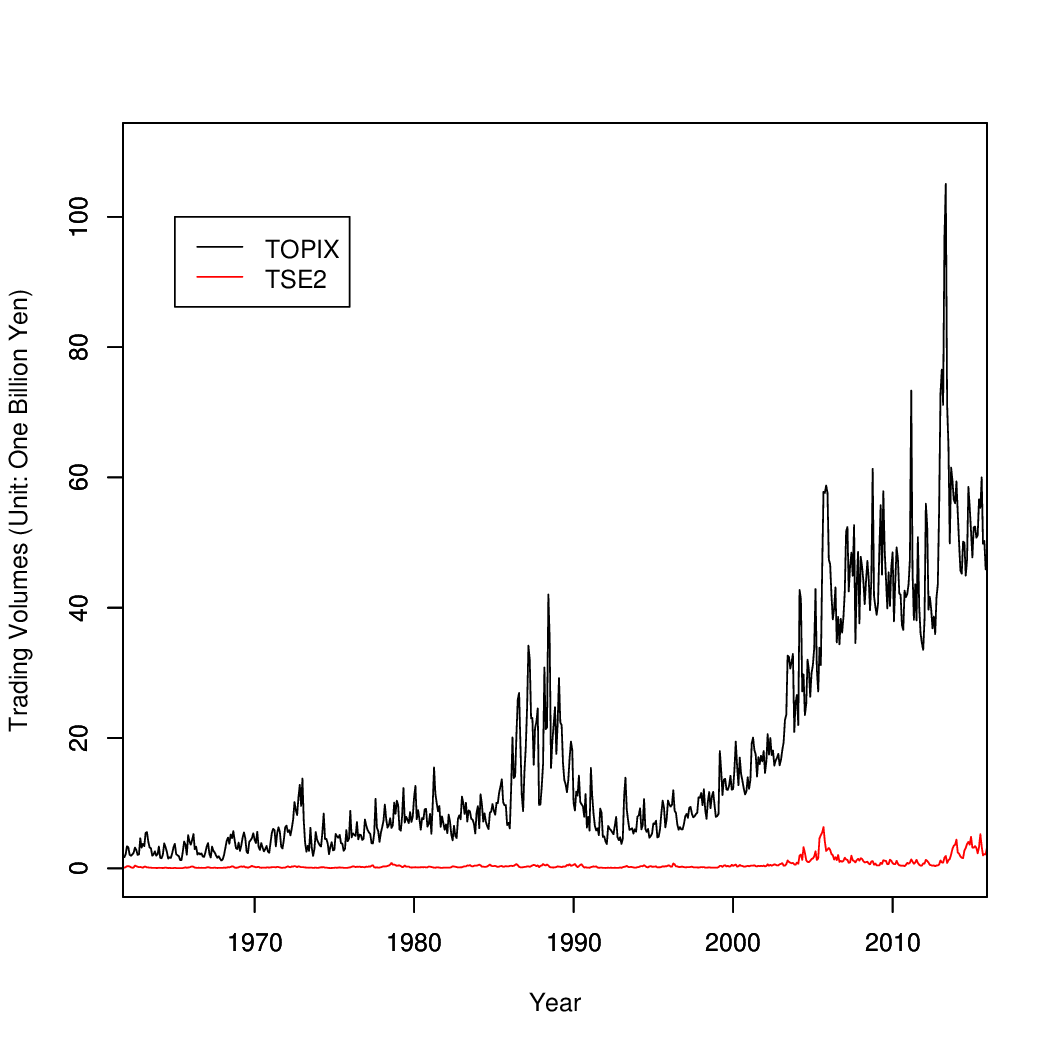}
 \includegraphics[scale=0.4]{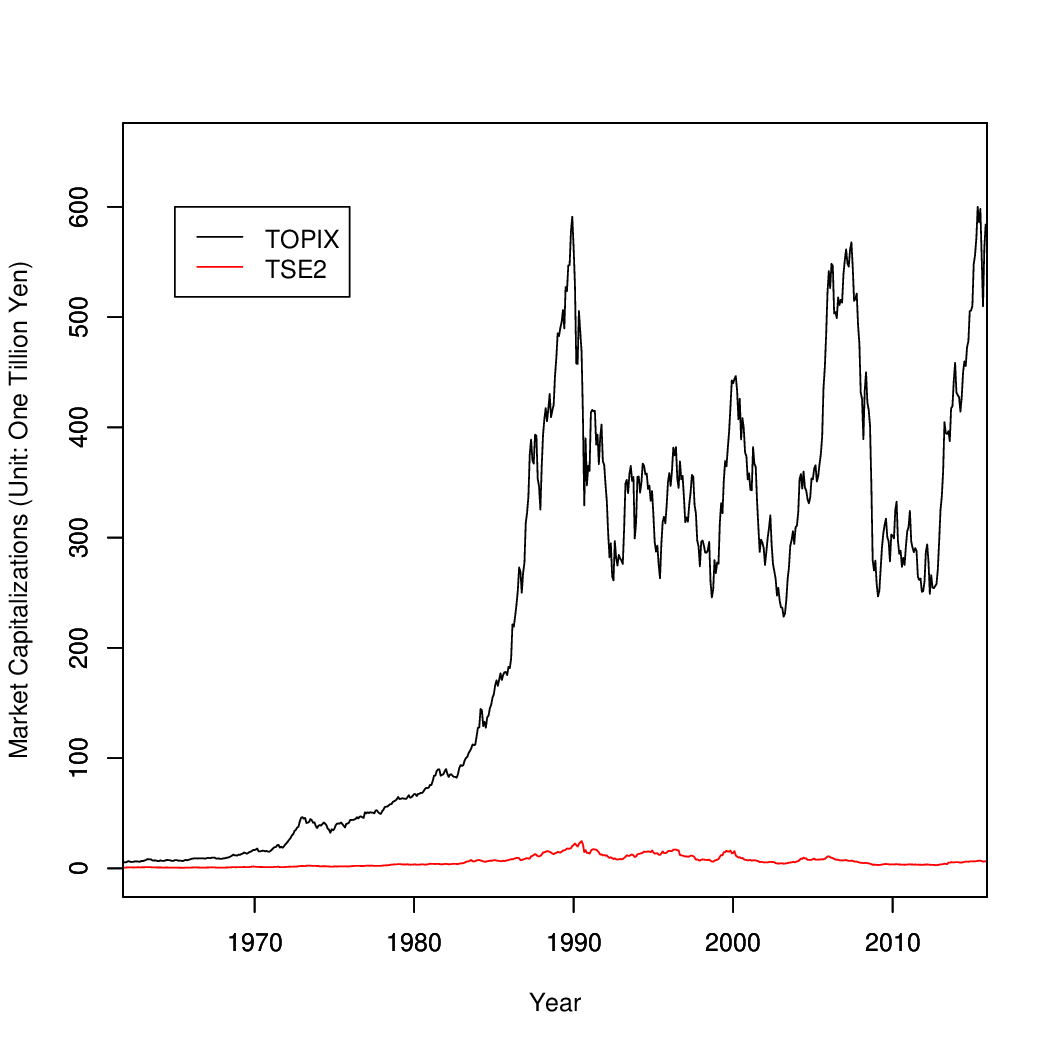}
\vspace*{5pt}
{
\begin{minipage}{420pt}
\scriptsize
\underline{Notes}:
\begin{itemize}
 \item[(1)] The panels of the figure show trading volumes (left panel) and market capitalizations (right panel) for the TOPIX and the TSE2.
 \item[(2)] The dataset is obtained from the web page of Japan Exchange Group (\url{http://www.jpx.co.jp/english/markets/statistics-equities/misc/index.html}).
 \item[(3)] R version 3.2.3 was used to compute the statistics.
\end{itemize}
\end{minipage}}%
\end{center}
\end{figure}

The different criteria for listing on the TOPIX and TSE2 in terms of the number of shareholders, tradable shares, and market capitalization of the shares listed might explain these differences in the Japanese stock market.
\begin{center}
(Figure \ref{amhjp_fig2} around here)
\end{center}
In particular, Figure \ref{amhjp_fig2} shows that trading volumes and market capitalizations are quite different between the TOPIX and TSE2. Those facts indicate that trade openness have been different between the two markets.\footnote{\citet{lim2011toi} show that trade openness is associated with stock market efficiency in 23 developing countries.} Figure \ref{amhjp_fig1} also shows that the degree of market efficiency of the TOPIX not only varies over time, but also has evolved since the bursting of the bubble economy in the early 1990s. The market efficiency of the TOPIX reflects the shock of the Asian financial crisis, Lehman Brothers bankruptcy, and monetary easing by the Bank of Japan since April 2013, whereas that of the TSE2 does not.\footnote{Our empirical result on the TOPIX is consistent with that presented in \citet{kim2008aas} who test the Martingale hypothesis in 1990s Asian stock markets by using the multiple variance ratio test.} Our empirical results thus support \citetapos{lo2004amh} AMH in the Japanese qualified stock market.

\section{Concluding Remarks}\label{sec:amhjp_cr}

This study examines \citetapos{lo2004amh} AMH in Japanese stock markets (TOPIX and TSE2). In particular, we measure the degree of market efficiency by using a time-varying model approach of \citet{ito2014ism,ito2016eme}, which provides a more accurate measurement of market efficiency than conventional statistical inferences (i.e., statistical tests using the moving window method). The empirical results show that (1) market efficiency changes over time in the TOPIX and TSE2, (2) the market efficiency of the TSE2 is lower than that of the TOPIX in most periods, and (3) the market efficiency of the TOPIX has evolved since the bursting of the bubble economy in the early 1990s, but that of the TSE2 has not. Therefore, we conclude that the empirical results support \citetapos{lo2004amh} AMH for the more qualified stock market in Japan.

\section*{Acknowledgments}

The author would like to thank the Editor, Brian Lucey, an anonymous referee, Mikio Ito, Jun Ma, Colin McKenzie, Taisuke Otsu, Tatsuma Wada, Toshiaki Watanabe, Tomoyoshi Yabu, Makoto Yano, and the seminar participants at Doshisha University and Keio University for their helpful comments and suggestions. The author would also like to thank the financial assistance provided by the Japan Society for the Promotion of Science Grant in Aid for Scientific Research Nos. 26380397 and 15K03542. All the data and programs used for this study are available upon request.


\bigskip

\bigskip

\bigskip

\setcounter{table}{0}
\renewcommand{\thetable}{\arabic{table}}

\begin{table}[!hbp]
\caption{Descriptive Statistics and Unit Root Tests}
\label{amhjp_table1}
\begin{center}
{\footnotesize
\begin{tabular}{cccccccccccccc} \hline\hline
& & & Mean & SD & Min & Max & & ADF-GLS & Lag & $\hat\phi$ & &
 $\mathcal{N}$ \\ \cline{4-7}\cline{9-11}\cline{13-13}
 & TOPIX & & $0.0043$ & $0.0442$ & $-0.2439$ & $0.1336$ & & $-18.7414$ & $0$
			     & $0.2969$ & & $650$ & \\
 & TSE2 & & $0.0058$ & $0.0524$ & $-0.2012$ & $0.1765$ & & $-14.7576$ & $0$
			     & $0.4968$ & & $650$ & \\\hline\hline
\end{tabular}}
\vspace*{5pt}
{
\begin{minipage}{420pt}
\scriptsize
{\underline{Notes:}}
\begin{itemize}
\item[(1)] ``ADF-GLS'' denotes the ADF-GLS test statistics, ``Lag'' denotes the lag order selected by the MBIC, and ``$\hat\phi$'' denotes the coefficients vector in the GLS detrended series (see Equation (6) in \citet{ng2001lls}).
\item[(2)] In computing the ADF-GLS test, a model with a time trend and a constant is assumed. The critical value at the 1\% significance level for the ADF-GLS test is $-3.42$'.
\item[(3)] ``$\mathcal{N}$'' denotes the number of observations.
\item[(4)] R version 3.2.3 was used to compute the statistics.
\end{itemize}
\end{minipage}}%
\end{center}%
\end{table}%

\begin{table}[!hbp]
\caption{Preliminary Estimations and Parameter Constancy Tests}
\label{amhjp_table2}
\begin{center}
{\footnotesize
\begin{tabular}{cccccccccccc} \hline\hline
 & & & $Constant$ & $R_{t-1}$ & & ${\bar R}^2$ & & $L_C$ & \\
\cline{4-5}\cline{7-7}\cline{9-9}
 & \multirow{2}*{$R_{TOPIX,t}$} & & $0.0030$ & $0.2978$ & & \multirow{2}*{0.0858} & & \multirow{2}*{38.0370} & \\
 & & & $[0.0017]$ & $[0.0428]$ & & & & & & \\ 
 & \multirow{2}*{$R_{TSE2,t}$} & & $0.0036$ & $0.3983$ & & \multirow{2}*{0.1562} & & \multirow{2}*{57.9233} & \\
 & & & $[0.0019]$ & $[0.0345]$ & & & & & & \\
\hline\hline
\end{tabular}}
\vspace*{5pt}
{
\begin{minipage}{420pt}
\scriptsize
{\underline{Notes:}}
 \begin{itemize}
  \item[(1)] ``$R_{t-1}$'', ``$\bar{R}^2$'', and ``$L_C$'' denote the AR(1)
	     estimate, the adjusted $R^2$, and the \citetapos{hansen1992a}
	     joint $L$ statistic with variance, respectively.
  \item[(2)] \citetapos{newey1987sps} robust standard errors are in brackets.
  \item[(3)] R version 3.2.3 was used to compute the estimates.
 \end{itemize}
\end{minipage}}%
\end{center}%
\end{table}%

\end{document}